\documentclass{eas}
\usepackage{graphicx}
\usepackage{xcolor}

\def\msun{\hbox{M$_\odot$}}

\def\t4{\hbox{t$_{\rm 4}$}}

\def\reff{\hbox{R$_{\rm eff}$}}
\def\mvbrightest{\hbox{M$_{\rm V}^{\rm brightest}$}}

\begin{document}

\title{Young Massive Clusters:  Their Population Properties, Formation and Evolution, and Their Relation to the Ancient Globular Clusters} 
\runningtitle{YMCs and Their Populations}
\author{Nate Bastian}\address{Astrophysics Research Institute, Liverpool John Moores University, 146 Brownlow Hill, Liverpool L3 5RF, UK}
%
%
\begin{abstract}
This review summarises the main properties of Young Massive Clusters (YMCs), including their population properties, particularly focusing on extragalactic cluster samples.  We discuss potential biases and caveats that can affect the construction of cluster samples and how incompleteness effects can result in erroneous conclusions regarding the long term survival of clusters.  In addition to the luminosity, mass and age distributions of the clusters, we discuss the size distribution and profile evolution of the clusters.  We also briefly discuss the stellar populations within YMCs.  The final part of the review focusses on the connections between YMCs and the ancient globular clusters, whether or not they are related objects and how we can use what we know about YMC formation and evolution to understand how GCs formed in the early universe and how they relate to galaxy formation/evolution.
\end{abstract}
\maketitle
\section{Introduction and Historical Development}


Traditionally, it was thought that open (OCs) and globular clusters (GCs) were two fundamentally different types of objects, with no overlap between the populations.  The classic view of GCs is that they are high mass, high density ancient systems that belong to the halo/bulge of the Galaxy.  OCs, on the other hand, were young, sparse, low mass systems that belonged to the disc of the Milky Way.  However, this clear dichotomy breaks down as soon as we look at our nearest galactic companions, the Large and Small Magellanic Clouds.  

These galaxies contain a large number of ``populous clusters" or ``young globular clusters", i.e. young (less than a few Gyr), dense and high mass ($>10^5$~\msun) clusters.  Examples include the $\sim100$~Myr  NGC~1850, the $\sim300$~Myr NGC~1856 and the $\sim1.5$~Gyr NGC~1846  clusters in the LMC, and the older Linsday 1 and NGC~339 in the SMC ($6-8$~Gyr, $2\times10^5$~\msun - e.g., Glatt et al.~2008).  These clusters have GC type masses and densities, but open cluster type ages.  

Schweizer~(1987) was amongst the first to propose that globular clusters were still forming in the local universe, and were not restricted to forming in the special conditions of the early universe.  He suggested that galaxy mergers were a natural place for GC formation, as it was thought that many elliptical galaxies were formed through the merging of spiral galaxies which caused extreme starbursts.  This field really began to take off in the early 90s, however, with the advent of the Hubble Space Telescope, which could peer into the crowded and complex environments of nearby galactic mergers and starbursts.  Holtzman et al.~(1992) used HST to study NGC~1275 and found a large population of ``bright blue clusters", with colours/magnitudes that suggested that they were young ($<100$~Myr old) and massive $>10^6$~\msun.  While such objects had been observed in another galaxy merger, NGC~7252 (Schweizer~1987) it took the exquisite resolution of HST to confirm that the objects were compact, $<15$~pc, i.e., true clusters.

After this initial discovery, a number of authors used HST to find and analyse huge populations of young clusters in the other mergers/starbursts such as the Antennae galaxies (Whitmore \& Schweizer~1995), NGC~7252 (Miller et al.~1997) and NGC~3256 (Zepf et al.~1999); i.e. across the full Toomre sequence (Toomre~1977).  These discoveries opened up an entire new field, that of the young globular clusters, or young massive clusters (YMCs - also known as super star clusters or SSCs).  

After their initial discovery in starbursts and mergers, subsequent works started finding massive clusters (and cluster populations) in nearby spirals (e.g., Larsen \& Richtler~2000) and dwarf galaxies (e.g., Billett et al.~2003).  It was soon realised that across a wide range of environments the cluster population properties were largely the same, however in starbursts one tended to find brighter, and presumably more massive clusters (e.g., Whitmore~2003).

With the ubiquity of YMCs in nearby environments, it was (and is) natural to view them as the young counterparts to the old GCs, which could offer us tremendous insight into how GCs formed and evolved, as we could watch their entire lifecycle play out within the nearest 10s of Mpc, instead of looking at redshifts greater than $\sim3$.  A number of theoretical works attempted to explicitly link what was known about YMC formation/evolution to that of GCs (e.g., Ashman \& Zepf~1992; Zepf \& Ashman~1993; Fall \& Zhang~2001).  

In turn, we came to view OCs as simply the lower mass end of  a continuous distribution of YMCs.  In fact, during this time, due in large part to the advent of efficient near-IR cameras, we began uncovering large populations of YMCs in the Milky Way, such as the Arches and Quintuplet clusters near the Galactic centre (Figer et al.~1999); Westerlund~1 (Clark et al.~2005), and the Red Super Giant Clusters 1 and 2 (Davies et al.~2007).  Hence, it appears that the observed dichotomy between the OCs and GCs in the Galaxy was largely caused by a selection effect, that since we live in the disc of a dusty spiral, we were simply unable to see many of the massive young clusters residing in the disc, whereas we have a largely complete view of the GC population in the bulge/halo.

In these lectures I will outline the general properties of YMCs and YMC populations.  In particular, I will discuss how we find YMCs in extragalactic surveys, what their mass/luminosity distributions look like, what their age distribution can tell us about their evolution, and will outline potential stumbling blocks along the way.  Additionally, I will discuss what is known regarding the stellar populations and stellar initial mass functions within them.  Finally, I will relate what is known about YMCs to what is and isn't understood about GCs, and how the former can be used to shed light on the latter.

Before beginning I would like to point out a number of recent reviews that cover various topics discussed in the current work: Portegies Zwart et al.~(2010) provide an overview of YMCs in Milky Way and nearby galaxies and discuss many of their population properties; Longmore et al.~(2014) discuss YMC formation and their early evolution; and Adamo \& Bastian~(2016) focus on YMC populations and how the host galactic environment affects YMC formation and evolution.  Finally, Kruijssen~(2014 - see also Brodie \& Strader~2006) discusses the links between YMCs and GCs.

\section{Cluster Population Properties}

\subsection{Cluster Selection}
\label{sec:selection}
When studying cluster populations the first step is to define a cluster sample.  Generally, when referring to ``clusters", authors adopt one of two meanings.  The first is that there is an over-density of stars, relative to the field, so that stars appear more ``clustered" than expected from a random/uniform distribution.  The second meaning is a gravitationally bound collection of stars.  In this second definition one normally has open clusters or globular clusters in mind, i.e., objects that have survived for many crossing times and hence are bound.  Unless one has access to the full 6 dimensional positional/kinematic information of all the stars in the region (along with a knowledge of the full gravitational potential of the region caused by the remaining gas and larger potential of the host galactic system) it is not possible to fully determine if a young system is gravitationally bound.

These two definitions are applied commonly in the literature, and have resulted in some confusion.  If one applies the first definition to star-forming regions, it appears that all stars form in clusters, i.e., nearly all stars form in a clustered fashion (e.g., Lada \& Lada~2003).  However, when using proxies to determine if a region is bound or not, i.e., the surface densities of regions, it appears that only a minority of stars form in bound clusters (e.g., Bressert et al.~2010).

In general, throughout this review, we will adopt the second definition, as we are interested in long lived clusters.  In catalogues of extragalactic cluster candidates, one can generally assume that objects that are centrally concentrated and have ages in excess of $\sim10$~Myr are clusters in this sense, as they have undergone multiple crossing times.  Candidates less than this age may be clusters, or they may be unbound associations (e.g., Gieles \& Portegies Zwart~2011).

An example of the impact of how young cluster candidates are selected and the resulting effects on the age distribution has been shown in Bastian et al.~(2012).  The authors found that a visual inspection of young cluster candidates ($<10$~Myr) that were selected by automated algorithms led to more than half being classified as unbound associations, rather than clusters.  M83 is at a distance of only $\sim4.5$~Mpc, so one HST WFC3 pixel corresponds to roughly 1~pc, allowing (at some degree of confidence) clusters to be differentiated from associations based on their morphology.  At larger distances (i.e., in the Antennae galaxies at $\sim20$~Mpc) such differentiations are not possible, so it is expected that at young ages, unbound associations dominate the cluster candidate catalogues.

{\em Hence, when analysing cluster populations properties that include young objects ($<10$~Myr), care must be taken with the interpretation, as many objects in the sample may be unbound.  When studying the age distribution of clusters candidates in extragalactic samples, age bins below $\sim10$~Myr should be treated as upper limits (as many of the objects may be unbound associations).}

One example that such ambiguities in terminology can be troubling is in our understanding of early cluster evolution.  Age distributions of ``clusters" often show a pronounced drop between ages of $<10$~Myr and from $\sim10-30$~Myr.  If this was the case for clusters, defined as gravitationally bound objects, then this would imply that some physical mechanism was necessary to invoke that would destroy substantial fractions of young clusters.  ``Infant mortality" or ``gas expulsion" has often been invoked to cause the rapid destruction of large numbers of clusters (see the recent review by Longmore et al.~2014).  However, if the ``cluster" sample included unbound associations (which is often the case for both Galactic and extragalactic systems) then this drop seen in the age distribution is simply a result of the unbound associations dissolving into the field, hence no new disruptive effect would need to be invoked.  Despite a large amount of theoretical work on gas expulsion in young clusters (e.g., see the review by Banerjee \& Kroupa~2016), it appears that current observations do not require such a mechanism (e.g., Longmore et al.~2014) as the differing definitions (among other reasons) of ``clusters" is largely to blame for the confusion.

\subsection{Deriving Cluster Properties}

The most common way to age date unresolved young ($<1$~Gyr) clusters is to compare their integrated light properties (i.e., colours/spectra) to simple stellar population models (SSPs).  This generally requires a few assumptions.  First, this assumes that clusters are well represented by SSPs, i.e., there is little or no age or metallicity spread within the clusters.  While many models for the origin of the multiple populations observed in globular clusters invoke large age spreads within young massive clusters, observations suggest that approximating them as SSPs is generally justified (e.g., Cabrera-Ziri et al.~2014; 2016a).  Additionally, one must assume a form of the underlying stellar initial mass function (IMF; a pure power-law IMF like a Salpeter~(1955) or one with a turn-over at low masses, e.g., Chabrier~(2003)).  

For low-mass clusters, the upper end of the IMF may not be well sampled, i.e. there may only be a handful of high-mass stars, whereas SSP models generally assume that the IMF is fully sampled.  This can cause cluster colours to vary in the regime of low numbers of stars for a given age/mass (e.g., Fouesneau \& Lan{\c c}on~2010; Silva-Villa \& Larsen~2011).  Due to the cluster initial mass function (discussed below), there are many more low-mass clusters in a population than high-mass clusters, hence deriving accurate properties for these clusters can be essential for many studies.  A number of recent studies have created SSP models with stochastically sampled stellar IMFs and compared these models with the observations in a Bayesian way to derive the most likely underlying age and mass distribution (e.g., de Meulenaer et al.~2013; Anders et al.~2013; Fouesneau et al.~2014; Krumholz et al.~2015).  Alternatively, if one restricts the analysis to the higher mass end of the cluster mass function ($>5000$~\msun) traditional fitting methods ($\chi^2$ fitting) can be used with some degree of confidence (e.g., Adamo et al.~2011).

Through comparison between the integrated colours/magnitudes of individual clusters, one can estimate the age, mass and extinction of each cluster in the sample.  However, not all colours are equally affected by age and/or extinction.  In order to break this common degeneracy, blue filters (i.e. U-band or bluer) as well as a number of optical filters (to sample the Balmer break) are required (e.g., Anders et al.~2004). Unfortunately, near-IR colours are generally not very sensitive to age (with one exception, see below), so this kind of game will become much more difficult in the era of JWST.  In addition to including broad band colours in the fit, an improvement that can be made is to include H$\alpha$ emission in fits (e.g., Chandar et al.~2010; Adamo et al.~2010) which can be used to differentiate an older low-extinction cluster from a younger ($<10$~Myr) more highly extincted cluster.  However, in cases where spatial resolution is limited (i.e. in distant sources) unrelated H$\alpha$ emission can enter the aperture, causing spurious results.  Hence, in these cases it is preferable to use a UV filter, which is also very sensitive to young ages, but directly traces the stars to make the differentiation (Hollyhead et al.~2016).

In addition to using the observed colours, other studies have compared integrated optical spectroscopy to SSP models to derive ages, which can in turn be used in combination with integrated colours to derive the extinction and age.  This has the benefit of largely removing the age/extinction degeneracy (e.g., Trancho et al.~2007).  Recent work has gone beyond using individual line indices and instead fits the full spectrum, with or without removing the continuum (e.g., Cabrera-Ziri et al.~2014; 2016a). 

If strong emission lines are seen in the spectrum (e.g., H$\alpha$, H$\beta$, $O[{\sc III}]$), with little or no absorption in the Balmer lines then the cluster is $<7-10$~Myr.  In addition to nebular emission lines, one can search for Wolf-Rayet (WR) features in the integrated optical spectrum.  This feature, at $\sim4650$\AA, is due to the winds of high mass evolved stars (WR stars) and their presence constrains the age of the cluster to $\sim3-7$~Myr (e.g., Sidoli et al.~2006).

\begin{figure}
\centering
\includegraphics[width=3.5cm]{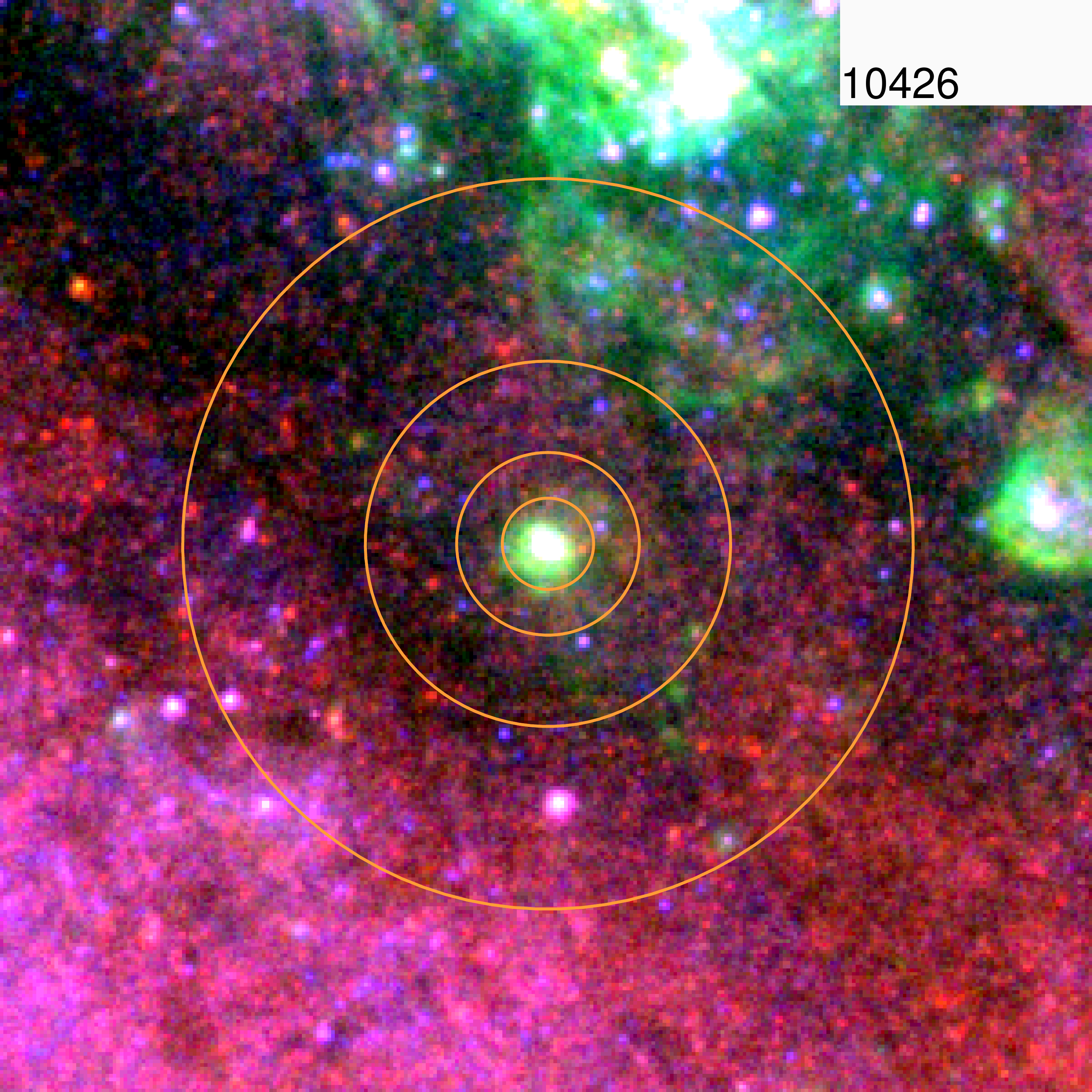}
\includegraphics[width=3.5cm]{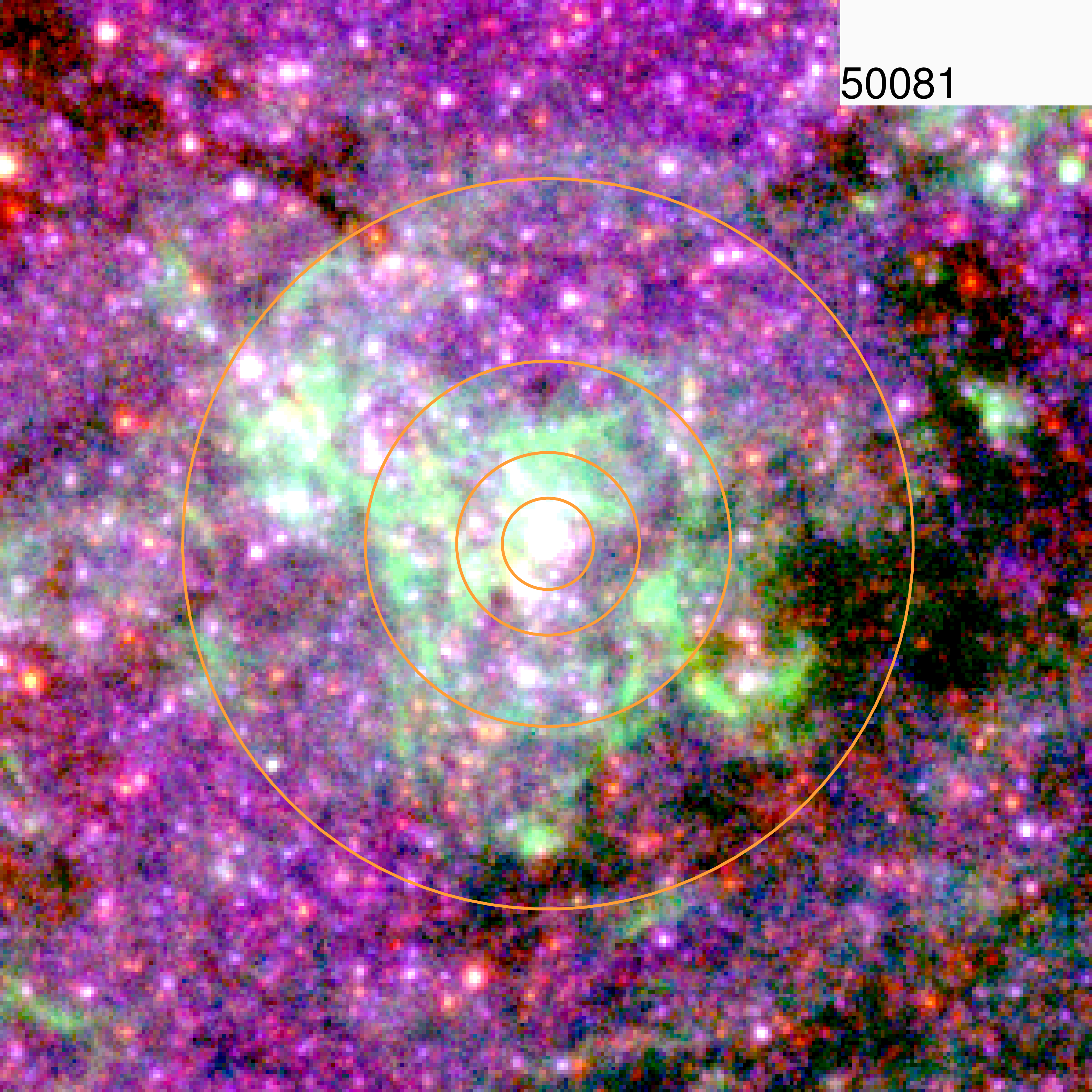}
\includegraphics[width=3.5cm]{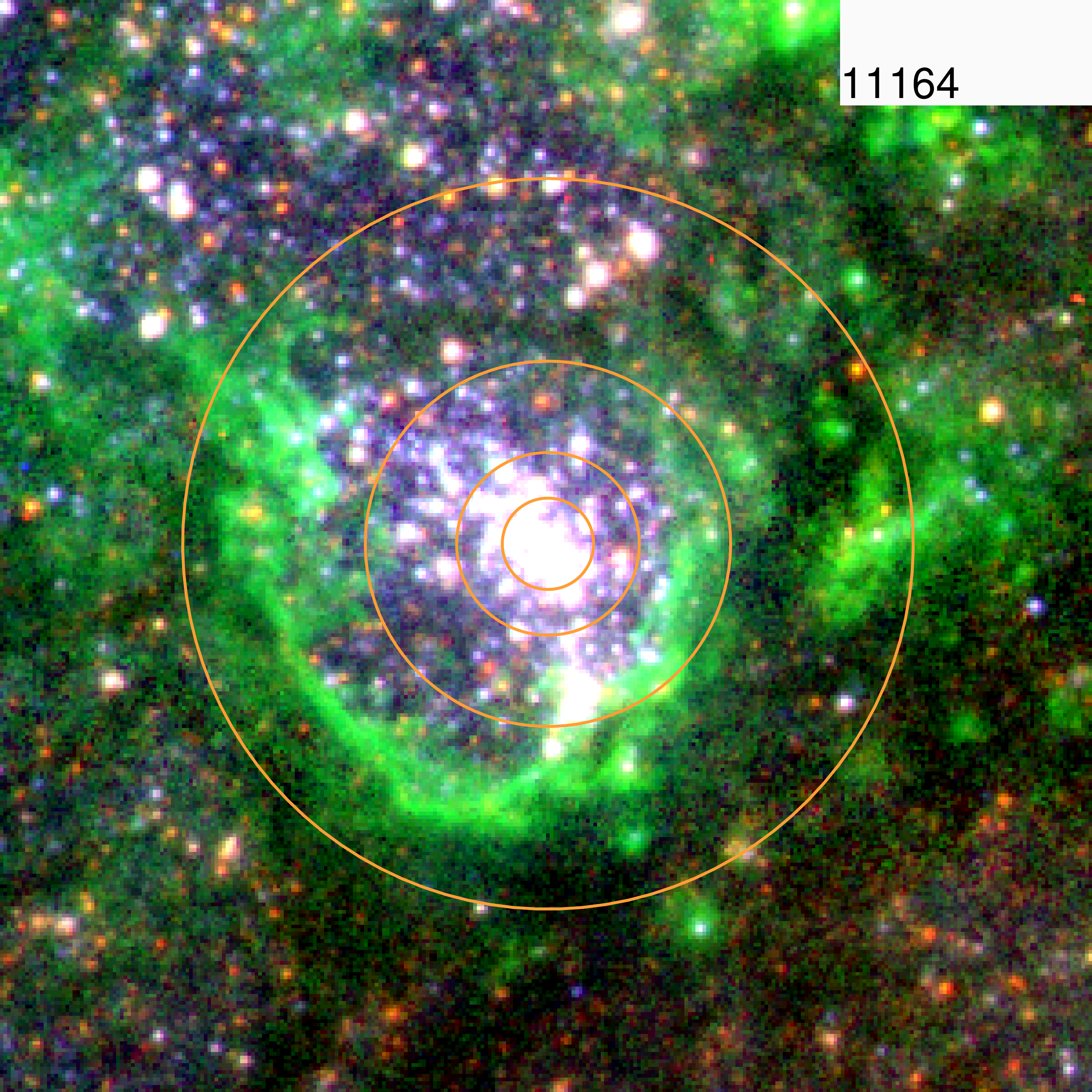}

\caption{Examples of YMCs in M83 as they evolve from being embedded (left panel), partially exposed (middle panel) to fully exposed (right panel), as can be seen by the clear ionised bubble surrounding the cluster.  This process takes $1-2$~Myr and appears largely independent of cluster mass (from $\sim 10^4 - 10^7$\msun). In each panel, blue represents the B-band, green traces H$\alpha$ emission (i.e., ionised gas) and red traces the I-band.  Figure taken from Hollyhead et al.~(2015). }
\label{fig:hollyhead}
\end{figure}

Whitmore et al.~(2011) suggested that the size of H$\alpha$\ bubbles around YMCs could be used as an age indicator, and compared the bubble size with the ages derived based on integrated colours.  Hollyhead et al.~(2015) expanded this idea and found that YMCs appear to be gas free (i.e. have bubbles surrounding them, but are no longer embedded in their natal giant molecular cloud) after 2-3~Myr (at least in M83, but also see Bastian et al.~(2014) for larger sample of massive, $>10^6$\msun, young clusters where this is also true).  An evolutionary sequence of YMCs from embedded to partially embeddded, to fully exposed is shown in Fig.~\ref{fig:hollyhead}.

Finally, Gazak et al.~(2013) found that the near-IR colours of a YMC is extremely sensitive to the presence of even a single red super giant.  They found, both through population models and in comparison with observations, that the near-IR colour of YMCs changes abruptly at the transition from having no RSGs to having just a few RSGs.  This transition appears to happen at $\sim6$~Myr, where the cluster goes from being extremely blue in the near-IR to being very red.  Extinction at young ages can complicate this method (as young embedded clusters can appear red) although this is less of a concern than in optical studies as it is based in the near-IR where the effects of extinction are lower.

\subsection{Selection Effects and Biases}
\label{sec:selection_effects}

Once the properties for individual clusters have been homogeneously derived, we can begin exploring the population properties of the clusters.  Of course, when doing so, one needs to carefully account for selection effects and biases.  Some effects are relatively obvious, i.e. if you are studying the luminosity function of clusters, the faint end of the LF will be more affected by completeness than the bright end, so any turn-over in the population should be treated with some skepticism (e.g., Whitmore et al.~2010).  Other effects can be more subtle so we need to keep in mind that when we look at a single cluster property (i.e. the age distribution) we are integrating over other properties (i.e., the mass and extinction distributions).

\subsubsection{The Effect of Incompleteness (Luminosity Limited Samples)}

One particularly important place where such complications can arise is when studying the age distribution of clusters.  Most observations are luminosity limited, meaning that some low-luminosity clusters are missed from the sample, whereas bright clusters are easily spotted, so they are included in the sample.  Due to stellar evolution, to first-order, clusters dim as a function of time in most photometric bands\footnote{One large exception to this is that from an age of $\sim6$~Myr Red Supergiants begin to evolve, which emit a substantial amount of flux in the red-optical and near-IR.  While the Red Supergiants are alive, the cluster can become more luminous with age in the red bands, and can actually be used as an age indicator (e.g., Gazak et al.~2013).}.  Due to this fading, it is easier, for a given mass, to detect younger rather than older clusters. 

\begin{figure}
\centering
\includegraphics[width=6cm]{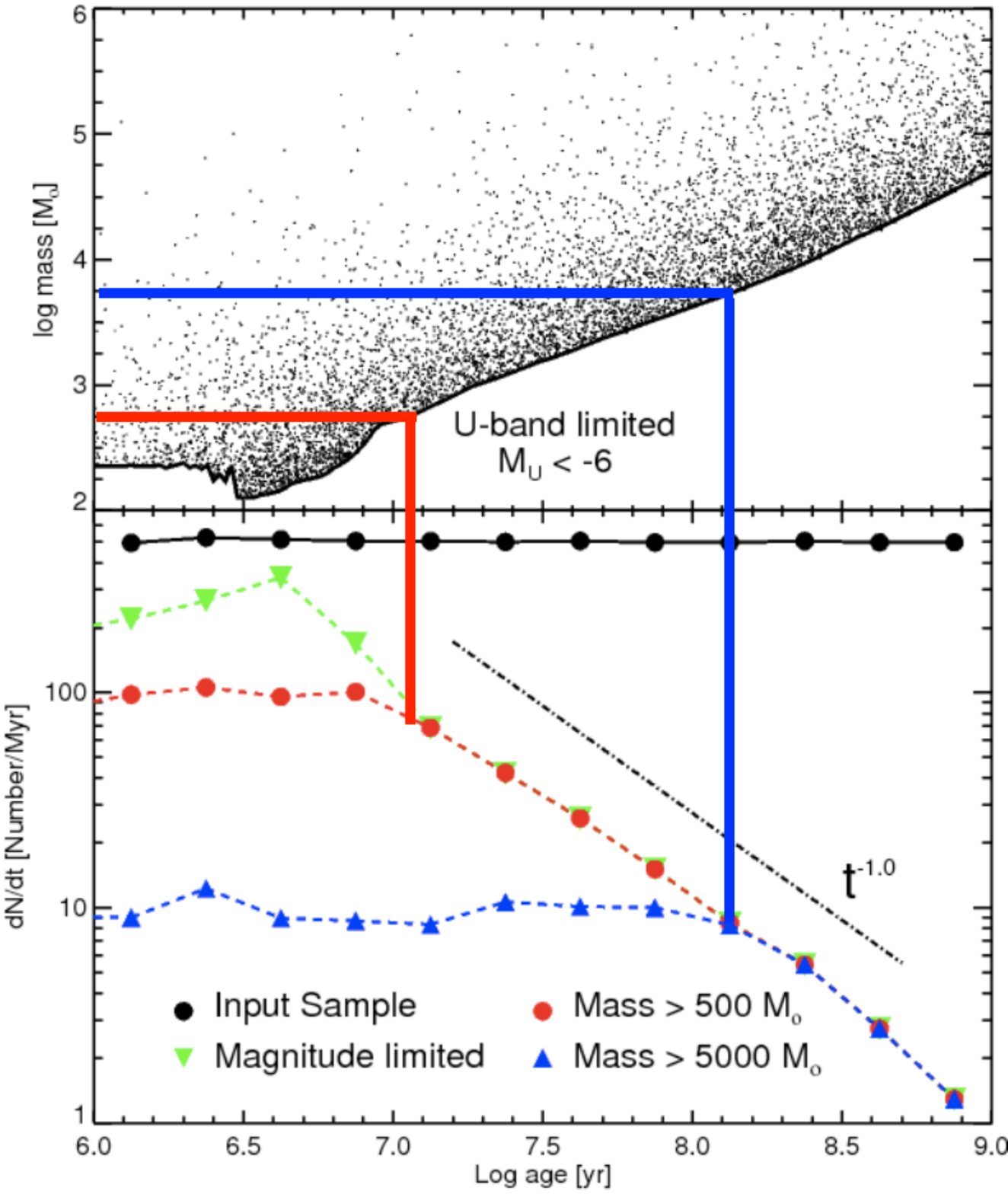}
\caption{A synthetic cluster population (constant cluster formation rate and normal cluster mass function) and the resulting age distributions for the 1) input distribution (black filled circles), luminosity limited sample (green upside-down triangles), and for two mass limits ($500$~\msun, filled red circles - and $5000$~\msun, blue triangles).  Note the strong effect that sample selection has on the conclusions.}
\label{fig:incompleteness}
\end{figure}

This can be most readily seen in the age-mass distribution of clusters in a population.  When looking at a figure of log(age) vs. log(mass), there is an ``incompleteness triangle", where there are few or no low mass old clusters in the sample (see the top panel of Fig.~\ref{fig:incompleteness}).  One of the main population properties that we wish to access is the age distribution of clusters (i.e. the number of clusters observed per linear age bin\footnote{One needs to correct for the logarithmic binning, i.e., that if the logarithm of age is used in the binning, each bins covers successively larger range of linear ages.}).  If one would just do this for the entire sample, this is effectively just integrating over mass for each age bin, and since samples extend to lower masses at young ages, this results in a declining age distribution for increasing age (i.e., it looks like all galaxies have been forming more and more clusters up to the present epoch).  Such an incomplete sample is known as ``luminosity limited".

To demonstrate the importance of this effect, and to visually see the best way to solve the problem, we create a synthetic cluster populations.  We assume a constant rate of cluster formation (500 clusters per Myr) for the past 1~Gyr (ages assigned stochastically), no cluster mass loss or dissolution, a lower mass limit of 100~\msun\ and no upper mass limit. Cluster masses are assigned stochastically, according to a power-law mass function, $Ndm \propto m^{-\alpha}dm$, with $\alpha=2$.  For each cluster we then have a mass and an age and we convert these into observational space (colours and magnitudes) using the simple stellar population models of Bruzual \& Charlot~(2003) at solar metallicity.  Finally, we ``observe" our population by applying a luminosity limit, assuming that the observations are limited in the U-band at $M_U = 6$.  

The results are shown in Fig.~\ref{fig:incompleteness}.  In the top panel, each dot represents a cluster and the effect of incompleteness is shown as a solid black line (this is simply set by the evolutionary fading of clusters below the detection limit).  In the bottom panel, we show the cluster age distribution for different sub-samples.  The black filled circles show the input sample.  Note that it is flat, since the input was a constant cluster formation rate. The green upside-down triangles show the resulting age distribution once the luminosity limit is applied.  This is what would be seen if the observed sample were used to construct the age distribution.  Note that it decreases rapidly, with a slope of $\sim-1$ (representing $dN/dt \propto t^{-1}$).  The rate of decline is filter dependent (since clusters fade at different rates in different filters), with the U-band or V-band limited samples declining at $t^{-1}$ or $t^{-0.6}$, respectively (e.g., Gieles~2010).

Finally, we show the results if two different mass limits (only looking at clusters more massive than a given limit) are applied, one at $500$~\msun\ (red filled circles) and $5000$~\msun\ (blue triangles).  In the lower mass limit case, we see that we retrieve the shape of the input age distribution (i.e., flat) until an age of $\sim10$~Myr, at which time the sample begins to drop due to incompleteness.  For the higher mass cut, the sample stays flat until $\sim100$~Myr, before dropping due to incompleteness.  Hence, the easy way to test if a sample is luminosity or mass limited, is to take progressively higher mass cuts.  If the shape (i.e., slope) of the age distribution becomes shallower for higher mass cuts, the sample is likely to be luminosity limited.

If incompleteness is not taken into account, which has happened in the literature, in order to explain the resulting age distribution of clusters, authors have adopted heavy cluster dissolution or mass loss (i.e., that say $\sim90$\% of clusters are destroyed each decade in age).  Hence, if cluster disruption is under discussion, samples should be mass-limited instead of luminosity limited (or if luminosity limited samples are used, care should be taken).

Note that mass limited samples can recover the underlying shape of the age distribution, but the total number of clusters is heavily affected, simply because the lower mass clusters are not included.  For example, the input age distribution and output age distribution with a $5000$~\msun\ limit are offset by a factor of 50.  This is because the lower mass limit between the two samples is different by a factor of 50 ($100$ vs. $5000$~\msun), which is a nice characteristic of a power-law mass distribution with an index of -2.  This needs to be taken into account if absolute numbers of clusters are needed, as is the case when studying the fraction of stars that form in clusters, which will be discussed in \S~\ref{sec:gamma}. 

\subsection{Size-of-sample Effects}
\label{sec:size_of_sample}

To see a key feature of cluster populations, we can create simple synthetic populations where we adopt a constant cluster formation rate\footnote{A constant cluster formation rate, on average, can be achieved by assigning clusters a random age over the time span of interest (e.g., for the past 1~Gyr).} and each cluster is assigned a mass, stochastically, from a power-law mass function with index $-2$ between some minimum (i.e., $100$~\msun) and maximum ($>10^7$~\msun) mass.  

\begin{figure}
\centering
\includegraphics[width=6cm]{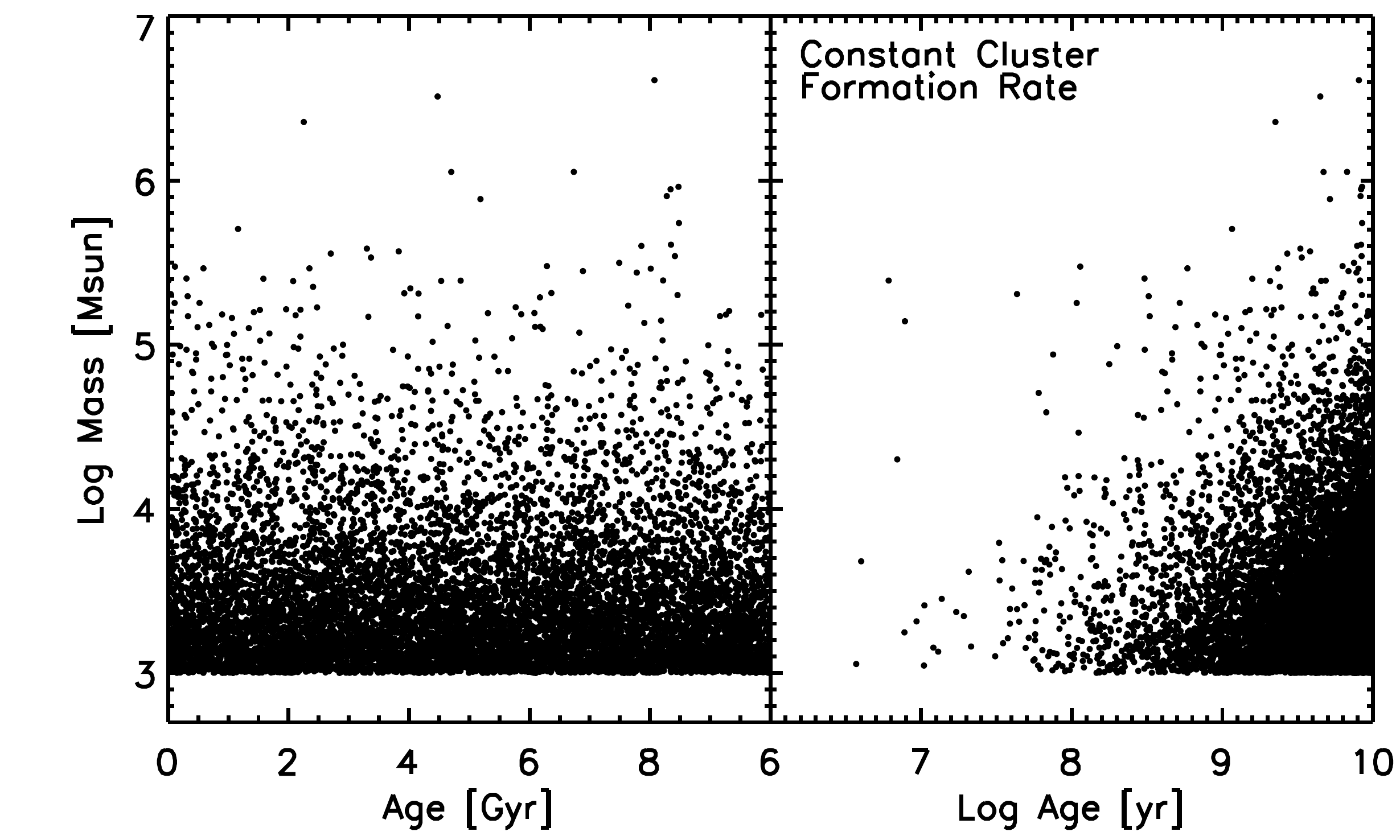}
\caption{Different representations of the same synthetic cluster population (constant cluster formation rate, power-law MF).  On the left the ages are plotted in linear units, whereas in the right panel they are plotted in logarithmic units.  Note that in the right panel it appears that there are more older clusters, but this is simply due to the plotting choice and is not physical.  Also note the increasing envelope of cluster mass as a function of increasing age in the right hand panel.  This shows a size-of-sample effect. (Made after Fig.~1 of Gieles et al.~(2006a))}
\label{fig:sos}
\end{figure}

One such population is shown in Fig.~\ref{fig:sos}. In the left panel, we show a synthetic cluster population (constant cluster formation rate, power-law mass function) with the x-axis (cluster age) plotted linearly.  The same population is shown in the right panel but now with the x-axis plotted logarithmically.   In the right panel the most-massive cluster as a function of time increases, i.e. the upper envelope of the distribution increases in mass as a function of time.  It appears that there are more older clusters, but this is only due to the logarithmic plotting.  As each bin in logarithmic age contains more and more clusters, the MF is sampled to higher masses, so the upper envelope of the masses increases as a function of logarithmic age. This is a size-of-sample effect, i.e. the more clusters that you have in a population, the more the underlying distributions (in this case mass) are sampled.  This effect impacts a number of population properties.

For example, Hunter et al.~(2003) used the upper envelope increase in the log age vs. log mass diagram to constrain the index of the mass distribution of clusters in the Magellanic Clouds, and found it to be slightly steeper than the nominal $-2$ index typically found for cluster populations.  Gieles \& Bastian~(2008) extended this analysis and showed that the increase of the upper envelope is controlled by the index of the mass function and cluster dissolution.  Their analysis showed that the high mass clusters in a sample of seven galaxies are not rapidly disrupted.  

Whitmore~(2003) showed that the strong relation between the number of clusters in a given population (brighter than a given magnitude limit) and the magnitude of the brightest cluster can be well explained by a size-of-sample effect, and like Hunter et al.~(2003), found that the implied underlying mass function of the clusters was steeper than the nominal $-2$.

An additional feature of the size-of-sample effect is that galaxies with higher star-formation rates (SFRs) form more clusters, which sample the underlying mass function to higher values.  This results in a  relation between the SFR of the host galaxy and the magnitude of the brightest cluster within the sample (e.g. Larsen 2002).  By building synthetic cluster populations, like that of Fig.~\ref{fig:incompleteness}, Bastian~(2008) showed that the observed SFR vs. \mvbrightest\ relation could be reproduced if $\sim8$\% of stars are formed in clusters that survive for more than $5-10$~Myr.  Additionally, they found a slightly steeper mass function than the nominal $-2$, something that all studies focussing on the brightest (most massive) clusters seem to find.  We will return to this point in Section~\ref{sec:mass_distribution}.

\subsection{Mass Distributions}
\label{sec:mass_distribution}
The distribution of cluster luminosities and masses has recently been reviewed at length in Adamo \& Bastian (2016).  To briefly summarise, the mass functions of clusters can be well described, over most of the mass range, as a power-law $Ndm \propto m^{-\alpha}dm$ with $\alpha = 2.0\pm0.3$.  This extends continuously from a few hundred solar masses (or less), as observed in the Milky Way (Piskunov et al.~2006) and nearby galaxies such as the LMC (Baumgardt et al.~2013), the SMC (de Grijs \& Anders~2006) and M31 (Fouesneau et al.~2014),  up to $\sim10^5$ or $10^6$~\msun\ in nearby  spirals (Chandar et al.~2010; Konstantopoulos et al.~2013) and starburst galaxies (Whitmore et al.~2010).

At the high mass end, many cluster populations display a truncation in the mass functions, at a mass that depends on their host environment (Gieles~2009; Larsen~2009; Bastian et al.~2012).  The cause of the truncation is likely to be found in the ISM properties of the host environment (e.g., Kruijssen~2014). 

\begin{figure}
\centering
\includegraphics[width=6cm]{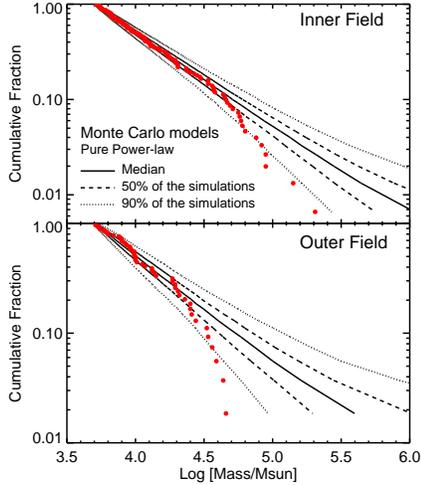}
\caption{The distribution of cluster masses with ages between 3 and 100 Myr in two fields within M83. Filled (red) circles represent the observations for each field. The lines denote the results of a series of Monte Carlo simulations that adopt the same number of clusters as the observations (only those used in the figure) for a pure power-law mass distribution with an index of $-2$ and no upper mass truncation. The solid line shows the median mass expected, while the dashed and dotted lines enclose 50\% and 90\% of the simulations, respectively. Note that both the inner and outer field distributions are inconsistent with the pure power-law case and require a truncation in the mass function at high masses.  Figure taken from Bastian et al.~(2012).}
\label{fig:mass_func}
\end{figure}

 In Fig.~\ref{fig:mass_func} we show the cluster MFs (for ages between $3-100$~Myr) for two different fields in M83 (Bastian et al. 2012), an inner field near the galaxy centre and an outer field adjacent to this further radially away (these fields correspond to Fields 1 and 2 in Silva-Villa et al.~2014).  Both populations show a power-law type behaviour at low masses, but both also show a truncation at high masses, with the truncation mass being higher in the inner region than the outer region.  To highlight the statistical significance of the truncation, we also show the results from a series of Monte Carlo simulations with a pure powerlaw MF and the same number of clusters as the observations.  In both cases the lack of higher mass clusters appears to be real.  Adamo et al.~(2015) have further investigated this trend with M83 by including an additional 5 HST/WFC3 pointings to search for radial trends and confirm these results based on a much larger sample of clusters.  Freeman et al. (in prep.) have investigated the link between the truncation mass in clusters and that found for GMCs based on ALMA observations, and find that the GMC mass function also displays a truncation.  They find that the most massive GMC is about a factor of 100 more massive than the most massive cluster within the same radial bins.

It is worth noting that techniques to measure the MF index using the high mass end of the distribution tend to derive indices steeper than studies that use the main body of the MF (i.e. measure the MF directly through a histogram).  This is consistent with the idea of a truncation or steepening at the high mass end.

Some theories for the destruction of young clusters due to gas expulsion have predicted that the MF of clusters should display clear features. For example, clusters below a given mass may be unlikely to form high mass stars, and hence would not suffer from the rapid gas expulsion hypothesised to fully destroy clusters (e.g., Boily \& Kroupa~2003).   Boily and Kroupa~(2003) predicted a peak around $10^3$~\msun\ in the MF of young clusters due to this effect.  Current observations are clearly at odds with this prediction.  The most detailed work to date on this has been done by Fouesneau et al.~(2014) on the M31 cluster population, using stochastically sampled SSP models, which are essential to use at these low masses.  They found a power-law MF with an index of $-2$ down to $\sim500$~\msun, with no evidence for the strong features predicted by Boily \& Kroupa~(2003).

This is another argument, in addition to those reviewed in Longmore et al.~(2014 - see also \S~\ref{sec:selection}), against the notion that gas expulsion and the subsequent dissolution is a key feature of young cluster populations.

\subsection{Age Distributions and Cluster Dissolution}

As with the mass distribution of young clusters, the age distribution and cluster dissolution has recently been extensively reviewed by Adamo \& Bastian~(2016).  Here, we will only summarise the main conclusions/context from that work.

The age distribution of clusters carries a tremendous amount of information about the population; its formation history (in principle over the entire galactic lifetime) as well as the rate of cluster dissolution over the same period.  Observations of cluster populations in post-starburst galaxies like NGC~7252 (e.g., Miller et al.~1997; Schweizer \& Seitzer~1998), NGC~1316 (Goudfrooij et al.~2004) or M82 (e.g., Konstantopoulos et al.~2009) clearly demonstrate that during major star-forming events, large populations of YMCs are formed.  Hence, it was clear, relatively early on in studies of YMCs, that they had great potential in determining the star-formation history of a galaxy.

However, an additional effect that can strongly alter the observed distribution of cluster ages is that of cluster dissolution, as clusters can lose stars to their surroundings due to two-body relaxation and gravitational interactions with the surrounding environment (e.g., galactic tidal field, GMCs, spiral arms, etc). Additionally, some clusters may become entirely destroyed due to these processes.

Theoretically, the evolution of a cluster, and its subsequent mass loss history, is relatively well understood.  Clusters with large numbers of stars (i.e., high mass clusters) should be less prone to two-body relaxation (i.e., it will take them much longer to dissolve than their lower-mass equivalents - e.g., Baumgardt \& Makino~2003), and dense clusters are less affected by e.g., GMC passages (e.g., Spitzer~1987; Gieles et al.~2006b).  Hence, the expectation is that more massive clusters should live for longer and that clusters in relatively weak tidal fields or in gas-poor regions should also live longer.

For a mass limited sample, we would then, for a constant cluster formation rate, expect the age distribution to be largely flat at young ages and then begin falling off at older ages as clusters lose mass and are disrupted.  However, observationally, there has been a lot of debate regarding the form of age distributions within galaxies.

Some works, e.g., Fall et al.~(2005); Whitmore et al.~(2007) and Chandar et al.~(2010) have argued for a quasi-Universal, environmentally independent age distribution, which is dominated by rapid cluster dissolution.  In this scenario, large fractions (up to $90$\%) of clusters are destroyed every decade in age, meaning that if there are 1000 YMCs with ages of $\sim10$~Myr, only 100 will survive to an age of 100~Myr, and only 10 clusters will survive to an age of 1~Gyr.  This empirical model was originally based on the Antennae galaxies, however it has also been suggested to apply to the Milky Way open clusters, the LMC and the SMC cluster populations (e.g., Whitmore et al.~2007).  However, applying the cluster population model of Whitmore et al.~(2007) to the Antennae galaxies, leads to a prediction that the current SFR should be above 250~\msun/yr, in clear contradiction to the observed $\sim20$~\msun/yr.  

Part of the reported rapid decrease appears to be due to sample selection, i.e. the inclusion of large fractions of associations in the youngest age bins (e.g., Bastian et al.~2012) and, in the case of the Antennae, the questionable assumption that the cluster formation rate has been constant.  The catalogue used by Whitmore et al.~(2007) for the Galactic open cluster distribution was not mass limited (it was distance limited which resulted effectively in a luminosity limited sample), which explains the observed rapid drop (see Fig.~\ref{fig:incompleteness}).  Mass limited studies of the open cluster age distribution show the expected flat distribution for the first few hundred Myr followed by a decrease due to disruption (e.g., Lamers et al.~2005; Piskunov et al.~2006).

Silva-Villa et al.~(2013) used a large survey of HST imaging of the nearby face-on spiral galaxy, M83, to study the age distribution of the cluster population as a whole, as well as within seven individual regions within the galaxy.  They found that the age distribution varied strongly between the different regions, being quite steep in the inner regions, where disruption is expected to be strongest, and becoming shallower at large galactocentric radii.  This is in good agreement with theoretical expectations and argues against a quasi-universal cluster age distribution.

Perhaps the best study to date has been carried out on the nearby galaxy M31, based on the Panchromatic Hubble Andromeda Treasury (PHAT) survey (Dalcanton et al.~2012; Johnson et al.~2015).  Fouesneau et al.~(2014) studied the age distribution of the clusters, where the ages, masses and extinctions had been estimated through a Bayesian analysis using stochastically sampled stellar IMFs, and found, as expected for this relatively weak tidal environment (and gas-poor), no evidence for a rapid decline in the number of clusters as a function of age.  They concluded that cluster disruption has not significantly affected the population of this galaxy.

Similar results, based on publicly available catalogues, although with traditional fitting methods (i.e., non-stochastically sampled stellar IMFs), have been found for the LMC (Baumgardt et al.~2013) and SMC (Gieles et al.~2007; de Grijs \& Goodwin~2008), where neither galaxy shows evidence for strong cluster mass loss.  This is expected given their relatively weak tidal fields and low gas densities.

While there has been quite a debate in the literature on the form of the observed age distribution of clusters, much of the debate was caused by differences in selecting the cluster sample, as well as the assumptions going into the analysis (i.e. a constant cluster formation rate for a starburst galaxy like the Antennae).  When well defined cluster samples are used, which are mass limited, and proper fitting of the resulting age distribution is undertaken, it is now clear that cluster disruption does not lead to a quasi-universal distribution, but is in fact strongly environmentally dependent.  This is in agreement with theoretical expectations of YMC populations and their evolution.

\subsection{Profile and Radius Distributions}

In addition to properties such as age and mass, another fundamental property of stellar clusters is their structure; how the stars are distributed within the host cluster.  While all clusters are centrally concentrated and the stellar density falls off with radius, they show relatively large variations in the basic structures, both in terms of their overall size and in how flat or steep their profiles are.  Old globular clusters are often observed to have a truncation at large radii, the tidal radius, which is thought to be due to the tidal field of the host galaxy.  A comparison between the profiles of YMCs and GCs, along with a general description of the types of profiles observed/fit, can be found in Schweizer~(2004).

In a pioneering study, Elson, Fall \& Freeman~(1987) found that the luminosity profiles of YMCs in the Large Magellanic Cloud (LMC) were best reproduced by a power-law profile/envelope, with no evidence of a turn-down or truncation at large radii.  They fit profiles of the type:  $I( r ) = I_0(1 + r^2/a^2)^{-\gamma/2}$, where $r$ is the distance from the cluster centre and $a$ is a characteristic radius.  This type of profile is known as an EFF profile.  One explanation for the differences observed between the old GC and YMC profiles is that all clusters may be born with power-law profiles which are subsequently truncated by the tidal field as the cluster evolves.  The authors found a  range in $\gamma$ values between $2.1$ and $3.35$.  In subsequent work authors have often defined $\alpha = \gamma/2$, for simplicity.  It is important to note that as one approaches or goes lower than $\alpha=1.0$ the effective radius of the clusters becomes undefined, as there is infinite area under the profile.  

Ma{\'{\i}}z-Apell{\'a}niz~(2001) studied a sample of 27 young clusters and associations in HST imaging and quantified their structures.  The author found that the sample could be split, roughly into two groups: young massive clusters and ``scaled OB-associations".  The difference between the two was that clusters had a centrally concentrated core whereas no such core was visible in the associations.  However, some clusters had extended halos which could mimic that of a surrounding associations, whereas others appeared to not have such a clear halo.  These results are similar to those discussed in \S~\ref{sec:selection}, highlighting the difficulty in defining a complete cluster sample at young ages, due to contamination of the sample due to associations.

The diversity in the profiles of YMCs can be seen by comparing those of NGC~3603 and R136, two clusters with similar ages,  masses and densities.  NGC~3603 appears to have a truncation in its profile at about $\sim1$~pc, at least in terms of its massive star content (Moffat et al.~1994).  Conversely, R136 has a relatively shallow power-law profile (even of its massive stars) that extends beyond 50~pc, and merges seamlessly into the surrounding association (e.g., Mackey \& Gilmore~2003; Campbell et al.~2010).  The youth of these clusters, and similarities of their other properties, suggest that their profiles reflect their formation and not any subsequent evolution.

Larsen~(2004) presented one of the first systematic studies of the structure of extragalactic YMCs, using a sample of clusters in 18 nearby spiral galaxies observed with the Hubble Space Telescope.  He measured the structure of the clusters using the custom code \texttt{ISHAPE} (Larsen~1999) which combines the camera point-spread function (PSF) with analytical cluster profiles and compares this to the observed clusters on the images.  For each cluster he could estimate the cluster's core radius and profile slope (i.e., a measure of how shallow or steeply a profile falls off with radius, i.e., $\alpha$ in the above definition), which can be combined to estimate the effective radius, \reff (the radius containing half the light of the cluster - in projection).  He found that the \reff\ distribution of the clusters has a peak near $\sim3$~pc, and a tail that extended to $>15$~pc. 

The cluster sample analysed in that work also showed a significant number of clusters with extremely shallow profiles ($\alpha < 1$), meaning that their effective radii were infinitely large.  Hence, if one wants to have an estimate of the effective radius, for example in order to estimate the dynamical mass of the cluster, one must artificially truncate the profile at a given, hopefully physically justified, radius (e.g., Larsen et al.~2001).

The author also noted an important problem with studying cluster sizes, that many of the very young clusters ($<10$~Myr) were located in crowded environments.  This means that due to contamination by the light of their neighbouring clusters, it can be difficult or impossible to accurately determine their structural parameters. 

Another result of Larsen~(2004) was that the clusters did not display a strong mass-radius relation, in particular that if such a relation existed it was much shallower than $M \propto R^{1/3}$.  This is important as it shows that higher mass clusters have, on average, larger densities.  Some disruption effects, like gravitational shocking from passing GMCs, are dependent on the cluster density, hence the observed distribution shows that such mechanisms should also be dependent on the cluster mass.

\begin{figure}
\centering
\includegraphics[width=6cm]{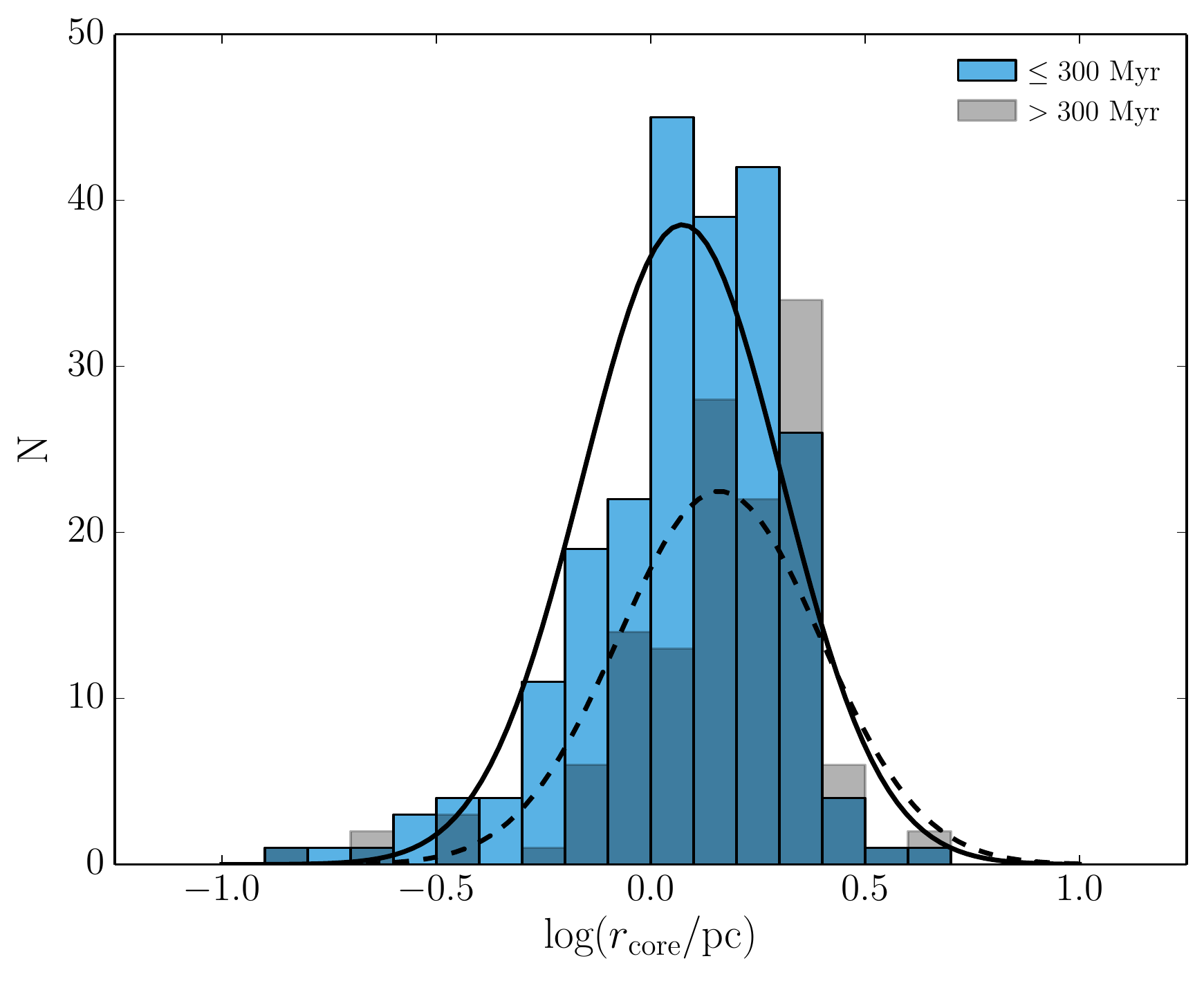}
\includegraphics[width=6cm]{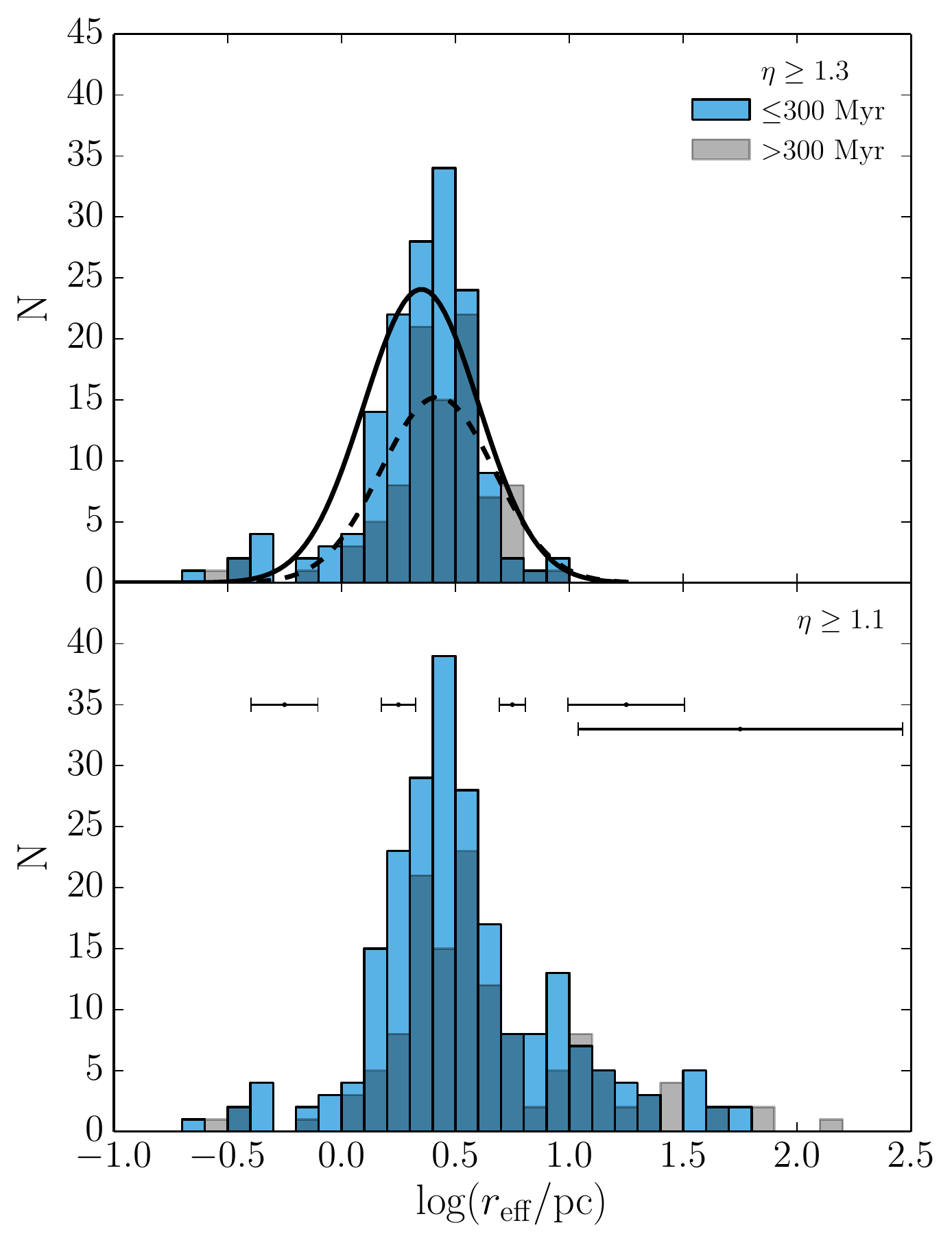}
\caption{{\bf Left panel:} The core radius distribution of clusters in the M83 sample found by Ryon et al.~(2015), limited to clusters with $\eta = \alpha > 1.1$.  The lines show the best Gaussian fits to the histogram.  {\bf Right panel:} The effective radius distribution for the sample.  The top panel shows the distribution found with clusters with well constrained profiles ($\eta = \alpha >1.3$ - i.e. the best fit sample with the lowest uncertainties) and the lower panel that includes some clusters with less well constrained profiles ($\eta = \alpha > 1.1$). }
\label{fig:sizes}
\end{figure}

Ryon et al.~(2015) have recently studied the structural properties of the cluster population of M83.  They also found a peak in the \reff\ distribution at $\sim2.5$~pc and a peak in the core radius distribution near $1.3$~pc.  In both distributions there was a trend observed with age, with young clusters having on average smaller core and effective radii.  The authors also attempted to find evidence for an evolution of the cluster profile, i.e. the turning of an initial power-law luminosity profile into a truncated King-type profile.  No clear evidence of a steepening of the profile with age or galactocentric distance was found.  Some of the distributions found by Ryon et al.~(2015) are shown in Fig.~\ref{fig:sizes}.

Ryon and collaborators also investigated how the effective radius of clusters varied as a function of other properties.  For example, they found that the mean \reff\ increased as a function of galactocentric distance, but that this was likely due to the intrinsic \reff\ vs. age relation, and the fact that the mean age of clusters increases as a function of distance from the galactic centre (due to cluster disruption being more efficient in the inner regions).  For younger clusters ($<100$~Myr), there did not appear to be a strong mass-radius relation in the sample.  This is in agreement with the earlier study of Larsen~(2004) and, as discussed above, has important implications for cluster disruption processes.  For older clusters, $100-300$~Myr, a mass-radius relation appears, consistent with a constant density (statistically, although the range of masses for a given radius was quite large).

The authors also estimated the Jacobi radii for their sample in order to estimate if the tidal field of the galaxy was influencing the cluster population.  They found that the population was mixed, with some clusters being tidally limited while others were under-filling and hence were not directly influenced by the tidal field.  However, GMC interactions can significantly affect clusters whose profiles are not tidally limited.  The authors conclude that mass loss due to stellar evolution and interactions between the clusters and passing GMCs (and the subsequent cluster expansion) were the dominant evolutionary processes affecting the population.

Bastian et al.~(2013c) studied the structural properties of some of the most massive YMCs known, focussing on the population of NGC~7252, which contains a number of clusters with masses well in excess of $10^6$~\msun, including one cluster, W3, with a mass near $10^8$~\msun\ (Maraston et al.~2004; Cabrera-Ziri et al.~2016a).  The clusters' profiles were best fit with EFF type profile, and two of the clusters (W3 and W30) had profiles extending beyond 300~pc from the cluster centre.  They also found a number of other massive clusters in the literature (see also Schweizer~2004) with very extended profiles, although most others could not be traced to such large distances.

\subsection{Cluster Complexes}

Young clusters are often not formed in isolation, but tend to form spatially (and temporally) near other clusters.  This is thought to be due to the hierarchical nature of the interstellar medium from which clusters form (e.g., Efremov \& Elmegreen~1998).   As a Galactic example, the Carina nebula regions contains a number of YMCs within it, along with a large field population, all within a $10-20$~pc region (e.g., Preibisch et 
al.~2011). 

The same applies even in starburst galaxies.  Zhang et al.~(2001) quantified the degree of association of young clusters using the auto-correlation function for a sample of YMCs in the Antennae galaxies.  Confirming what can be seen by eye on the images, they confirmed that the YMC spatial distribution was not random, but that YMCs tend to be found in regions containing other YMCs.  They applied their method to CO clouds (the progenitor sites of YMC formation) and found similar results.  They also looked at older cluster samples and found that the correlation dropped significantly with time, i.e. that clusters dispersed from their natal environment over timescales of 10s to 100s of Myr.

Bastian et al.~(2005) studied a sample of cluster complexes in M51, and found that unlike individual clusters, cluster complexes followed a clear mass-radius relation, going as $M \propto R^{0.5}$.  This relation is very similar to that found for GMCs in the same galaxy, suggesting that the complexes keep some of the properties of their host cloud, whereas the individual clusters have sizes that appear to be essentially independent of their mass.  This potentially could be a reflection of the chaotic birth environment within massive YMCs.

\subsection{The Fraction of Stars Forming in Clusters}
\label{sec:gamma}

As discussed in \S~\ref{sec:selection} the definition of ``cluster" can vary between authors and lead to very difficult conclusions if consistent terminology is not employed.  Another case in point is that of the fraction of stars that form in clusters.  If by ``cluster", an investigator is referring to an over-density of young stars relative to the field, then nearly $100$\% of stars form in clusters, i.e. very few stars form in total isolation.  This fraction is drastically reduced if one looks at gravitationally bound clusters, i.e. where associations have been specifically removed either by studying slightly older cluster populations ($>10$~Myr) or for younger populations based on their morphology.

The fraction of stars that form in clusters (known as $\Gamma$ - Bastian~2008), and its dependences on the ambient environment has recently been reviewed in detail in Adamo \& Bastian~(2016).  Here we will just summarise the main points.  For most quiescent spirals and star-forming dwarf galaxies, $\Gamma$ is between $3-15$\% (e.g., Ryon et al.~2014).  For starburst galaxies this can increase to nearly $50$\%.  However, this increase does not appear to be directly linked to the SFR of the host galaxies, but rather to the intensity of the starburst event, i.e. on the SFR surface density ($\Sigma_{\rm SFR}$ - the SFR per unit area), with more intense star-forming events forming a higher fraction of their stars in clusters (e.g., Larsen \& Richtler~2000; Adamo et al.~2011; Kruijssen~2012).  Due to the dependence on $\Sigma_{\rm SFR}$, $\Gamma$ also varies within galaxies, as seen by the relative steep drop from the centre of M83 ($\Gamma  \sim 26\%$) to the outer regions ($\Gamma \sim 8$\% - Adamo et al.~2015).

\section{Stellar Populations within YMCs}
\label{sec:stellar_pops}

The lectures in this volume by Estelle Moraux thoroughly cover what is known about the stellar populations within open clusters, focussing largely on the form of the stellar initial mass function (IMF) within them.  We also point the interested reader to the recent review on the IMF by Offner et al.~(2014) which covers many of the observational constraints in clusters as well as a review of the theories for the origin of the IMF. One of the big questions about YMCs, when they were first detected and studied in detail, is do they have the same kind of IMF, or is the star-formation process somehow different in these extreme environments?

First, it is worth stating that the present day mass function within GCs, which due to stellar evolution over the past 10-13 Gyr, only extends up to $\sim0.8$~\msun, is consistent with the IMF inferred for the field and the young lower mass/density open clusters, once the effects of dynamical evolution are taken into account.  The interested reader is referred to the reviews of Chabrier~(2003) and Bastian et al.~(2010) for an extended discussion of the IMF and present day stellar MF of GCs.  

For a handful of YMCs in the Galaxy or nearby galaxies, it is possible to resolve the clusters into the constituent stars and (more or less directly) infer the present day mass function (PDMF) of the cluster.  If one assumes little dynamical evolution of the cluster, the PDMF can be taken as the IMF, modulo stellar evolution at the high mass end.   While, to date, the IMF has not been studied down to sub-solar masses, the upper end of the IMF is generally well approximated by a single power-law, with a slope similar to, or shallower than a Salpeter~(1955) index (e.g., Kim et al.~2006; Andersen et al.~2009).  Hence, like in less extreme environments, there are many more lower-mass stars than higher mass stars.

However, many resolved studies have also found evidence that the measured index of the high-mass portion of the IMF is shallower than expected from a Salpeter type distribution.  Most results have found that in the inner regions of clusters the IMF is shallower than in the outer region.  This is known as mass segregation, i.e. a preference for the high mass stars to be located in the centre of the cluster.  Due to observational effects, great care must be taken in interpreting these results, as most selection effects push the observer to measure a shallower slope in crowded regions than in less crowded regions (e.g., Ascenso et al.~2009).  For very young clusters, that still contain significant structure in the spatial distribution of stars, mass segregation is even harder to quantify (e.g., Parker  \& Goodwin~2015).  However, in many cases such segregation can be taken to be reasonably certain (e.g., for Westerlund 1 -  Gennaro et al.2011).

If such segregation was primordial in nature, i.e. if the cluster was born with a radially varying IMF, this would be a clear indication of the non-universality of the IMF, implying that the environment does affect the type of stars that are born in a given volume, at a given time.  However, even for very young clusters, such mass segregation, can in principle be due to dynamical processes, namely the merging of sub-clumps (e.g., McMillan et al.~2007; Parker et al.~2014).

We can also get a somewhat cruder measure of the IMF in more massive extragalactic clusters by estimating their dynamical masses, through measuring their effective radii and velocity dispersions and applying the Virial theorem.  We can then compare the measured mass-to-light ratio (M/L) of the clusters to predictions from simple stellar population models for the appropriate age.  One of the important ingredients to the SSP models is the underlying stellar IMF, so we can vary that until we get a good agreement between the observed and modelled cluster.  This has been done for approximately $\sim20$ YMCs (ages between 5~Myr and 3~Gyr - see the compilation by Bastian et al.~2006).  Younger clusters ($<20$~Myr) tend to have higher M/L ratios than model predictions, although this is likely due to the fact that high mass stellar binaries tend to artificially inflate the inferred velocity dispersion (Gieles et al.~2010).  However, with one exception (M82-F, see Smith \& Gallagher~2001 and Bastian et al.~2010 for a detailed discussion of this cluster) all the other clusters are consistent with a Chabrier or Salpeter type IMF.  However, large variations in the underlying stellar IMF are not consistent with the data.

Hence, it is still unclear whether the stellar IMF does vary systematically as a function of environment.  However, extreme variations, as had been previously suggested (i.e. no stars in a population below say 2 or 3~\msun) can confidently be excluded.  More subtle variations may still be hiding due to the relatively large error bars associated with these studies.


\section{Linking YMCs to the Ancient Globular Clusters}

Ever since the discovery of young clusters with masses and densities that are similar to, or even exceed, those of the ancient globular clusters, there has been debate about whether YMCs are merely a younger version of GCs or if the ancient clusters are fundamentally different.  If YMCs are simply GCs forming today, then they can offer unprecedented insight into GC formation and evolution, due to their proximity.  In this perspective, the differences observed between YMCs and GCs merely reflect a Hubble Time's worth of evolution, both in the cluster's properties (i.e., mass and density) but also in their location relative to their host galaxies.

\subsection{Mass Functions and Migration}

One of the clear differences between YMC and GC populations is their underlying mass functions.  YMCs have a power-law type distribution that continues from the completeness limit (usually between $10^3 - 10^4$~\msun\ for most studies) to high masses (with a potential truncation at high mass that appears to be galaxy dependent - see Adamo \& Bastian~2016 for a detailed review).  GCs, on the other hand, have a log-normal distribution both in spirals like the Milky Way (e.g., Gnedin \& Ostriker~1997) to massive early type galaxies like those in the Virgo cluster (e.g., J\'ordan et al.~2007).  If GCs are nothing but the evolved cousins of YMCs, this implies that the majority of the initial clusters have been destroyed, primarily the lower-mass clusters (e.g., Elmegreen \& Efremov~1997).  There have been many claims in the literature for ``turn-overs" in a YMC mass function due to the preferential destruction of lower-mass clusters, however, it is a difficult measurement to make, as selection effects also can cause a turn-over (see Smith et al.~2007 for a demonstration of such a surface-brightness effect that can cause a turn-over).  Probably the most robust case for a turn-over in the MF of a young/intermediate age population is that of the 3~Gyr post-merger galaxy, NGC~1316, where Goudfrooij et al.~(2004) found a turn-over for clusters near the galactic centre whereas the population far from the centre of the galaxy was consistent with a pure power-law MF.

Fall \& Zhang~(2001) modelled the Galactic GC population, starting with a power-law (or Schechter) mass function and included the effects of two-body relaxation and the tidal effects of the host galaxy.  The authors were able to turn the original power-law to a log-normal distribution.  However in so doing, they had to make a series of assumptions that limit the applicability of the result. First, they assumed that the current potential of the Milky Way has always been as it appears today, i.e., they did not include the hierarchical buildup of the galaxy.  Secondly, in order to make the mass-loss from tidal effects strong enough they needed to put many clusters on highly elliptical orbits, essentially giving most clusters the same peri-galactic distance.  Their model also led to a prediction that the peak of the mass function should vary with Galactocentric radius, with the peak shifting to lower masses at further distances.  This is at odds with observations that shows a near universal peak (e.g., J\'ordan et al.~2007).

Recently,  Gieles \& Alexander~(2015) have expanded on this type of model, using the fast GC population code \texttt{EMACSS} (Evolve Me A Cluster of StarS).  They used the current mass distribution of GCs and inverted the problem, to infer the most likely form of the initial mass function of GCs.  The code includes relaxation, stellar evolution, tidal effects and changes in the cluster size with time.  The authors conclude that the initial mass function of GCs was not similar to that seen for YMCs, but must have been significantly shallower, and that it was not possible to end up with a near-universal mass function of GCs if they began with a power-law type distribution.  One effect that was not included in the simulations was that of cluster mass loss due to interactions between the galactic ISM (especially in the gas-rich birth environment of GCs) and the clusters.

In the discs of spiral galaxies today, the largest killer of clusters is not the tidal field of the host galaxy, but is rather ``tidal shocking" between GMCs and young clusters (e.g., Gieles et al.~2006b).  GCs likely did not form in their current positions in the halo, but more likely formed in the discs of gas rich proto-galaxies and subsequently migrated to their current positions due to galaxy interactions and mergers.  This effect was first explored in Elmegreen~(2010).  The author found that in the high gas density/pressure environment where GCs are thought to have formed, that indeed interactions with the surrounding gas could disrupt a significant fraction of clusters on short timescales (tens of Myr).  This effect then could remove large fractions of the low mass clusters of an initial population.  If, however, the cluster was able to migrate to an area of lower gas density/pressure the proto-GC would stand a much higher chance of survival (e.g., Kruijssen et al.~2011).  Since, all GCs likely formed in similar gas rich environments, it is thought that such rapid cluster destruction could lead to a largely universal, log-normal type mass distribution, i.e. similar for spirals, dwarfs and early type galaxies.

This idea has recently been further developed by Kruijssen~(2015) who used merger trees from cosmological dark matter only simulations with recipes for the locations and properties of proto-GC formation.  The author found that the vast majority of mass loss within a cluster population (and hence cluster dissolution) was during this early gas rich phase, and that the subsequent evolution of the surviving GCs (i.e. those that were able to migrate to the halos of their host galaxies) was only a minor component of the cluster mass loss, i.e. that tidal effects and two-body relaxation were minor players in the destruction of clusters.

This idea is particularly exciting, as it can explain, rather naturally, the near-universal turn-over of the GC mass function, and why even low mass dwarfs do not contain many clusters below $10^5$~\msun. If tidal disruption was the dominant mass-loss mechanism then in low mass dwarfs, which have weak tidal fields, one would expect that their full cluster population (even the low mass clusters) would still be present.  However, with an early gas rich phase of cluster mass loss, in most environments that can form GCs, the majority of low mass clusters will be rapidly disrupted.  The next step in  exploring this idea is full simulations, including the formation and subsequent evolution of proto-GCs in full hydrodynamical simulations of galaxy formation and evolution.  While it remains unfeasible to resolve individual GC formation and follow their evolution through N-body simulations, it can be done in a sub-grid manner and calibrated by N-body models.  Currently, this is being explored by a number of groups, e.g., Pfeffer et al.; Kruijssen et al. and Renaud et al.~(in prep.).

One observation that is potentially at odds with this interpretation is that of the relative fraction of metal poor stars in the field and in clusters in the dwarf spheroidal Fornax galaxy.  Larsen et al.~(2012) found that below a metallicity of [Fe/H]$ = -2$, GCs contained between $20-25\%$ of all the stars.  As the five known GCs in Fornax are all relatively high mass ($\sim10^5\msun$ - with a distribution not unlike that of the MW GCs), then if the mass function (MF) of the young GCs was a pure power-law, then all of the low mass clusters must have been dissolved and should be located in the field.  For a power-law MF with an index of -2, and a lower mass limit of $100$~\msun, this means that three-quarters of all the stars initially within GCs have been returned to the field (assuming that all clusters with masses above $10^5$~\msun\ are still alive and have not lost significant fractions of their initial mass).  Hence, even under the extreme assumption that 100\% of all stars formed in clusters in this galaxy, the observations are in tension with predictions, potentially suggesting that the GCs in Fornax were not born with a power-law MF.  There are a number of uncertainties in the above calculations, and the predictions/observations can be uncertain to within a factor of 2 or 3, so there is still some room to make the GC distribution consistent with YMCs, but further observations (e.g., Larsen et al.~2014) of GCs in dwarfs is making this more and more difficult.

\subsection{Including GCs/YMCs in Galaxy Formation/Evolution Simulations}

As GCs are some of the oldest luminous structures in the Universe, their formation and evolution must be intricately linked to the formation and subsequent evolution of their host galaxies.  We refer the interested reader to the review of Brodie \& Strader~(2006) for an overview of the use of GCs to trace galaxy formation and evolution.

While it appears to be straightforward to associate the formation of a globular cluster population with the formation of the underlying stellar component of the galaxy, a number of works that have attempted to trace the formation and subsequent evolution of GCs within cosmological simulations have not been able to reproduce the observed properties (e.g., Beasely et al.~2002).  One of the recurring problems is the difficulty in reproducing the observed bi-modality metallicity distribution of the Milky Way globular cluster population (e.g., Harris~1996) without invoking an ad-hoc truncation in the epoch for the metal poor globular cluster formation.  Other problems with such attempts have recently been reviewed in Kruijssen~(2014).

Early work in including GCs in cosmological simulations often relied on dark matter only simulations, which included star-formation and cluster formation/evolution in a semi-analytic way.  As simulations improved (e.g., Kravtsov \& Gnedin~2005), they allowed one to follow the formation of a Milky Way type galaxy, including baryonic physics (resolving the location of where GCs form) but generally did not include cluster disruption effects.  Other advances in our understanding of the formation of  GC populations can be found in e.g., Muratov \& Gnedin~(2010) and Tonini~(2013; and reference therein).

But for the present review we focus on the inter-relation between GCs and YMCs.  In particular, we want to include what is now known about YMC formation, their scaling properties with the host galaxy environment, and their subsequent evolution, to be able to track where GCs formed and how they evolved.

On this side of the problem, there has been significant progress made in using GCs to trace the buildup of galaxies.  A recent example is that of Mistani et al.~(2016).  They used the Illustris cosmological simulations to study a sample of dwarf galaxies in a galaxy cluster and also in the field.   They selected a sample of dwarfs with the same mass in both environments, and found that while the galaxies had the same final stellar mass, their star-formation histories (SFH) differed considerably.  The average dwarf in the galaxy cluster formed its stellar mass early on through a starburst as it entered the galaxy cluster, whereas the field dwarf formed its stars in a more continuous fashion.  As discussed above, the intensity of how stars form can affect the type of clusters formed and even the fraction of stars that are forming in clusters.  In intense starbursts, a higher fraction of stars are formed in clusters, as well as leading to a better sampling of the underlying cluster mass function (i.e., more massive clusters are likely to be formed).  Hence, Mistani et al. conclude that dwarfs in clusters should have more GCs (i.e. higher mass surviving clusters) than dwarfs in the field.  This is exactly what is observed (e.g., Miller \& Lotz~2007).

Pfeffer et al. (in prep.) and Kruijssen et al. (in prep.) have included the formation, evolution and destruction of YMCs (i.e. proto-GCs) in the EAGLE hydrodynamic simulations of galaxy formation and evolution (Schaye et al.~2015).  The simulations include the effects of the ISM on the type of YMCs that form (i.e., in high $\Sigma_{\rm SFR}$ environments the cluster mass function is sampled to higher masses, and the truncation mass is linked to the Toomre mass of the host galaxy) as well as the effect of GMCs in destroying the young clusters, cluster migration out of the cluster-forming environment, two-body relaxation in the clusters as well as the effects of mass loss due to the tidal field.  Additionally, the fraction of stars that form in clusters is also explicitly tied to the conditions of the host galaxy, with low $\Sigma_{\rm SFR}$ dwarfs forming only a few percent of their stars in clusters to starbursts forming $\sim50\%$ of their stars in clusters.

The epoch of formation, metallicity, and location of each forming cluster ``particle"\footnote{The ``clusters" are treated as star particles in the simulations, and their evolution is tracked in a sub-grid manner.} is required, and their location is tracked in time.  By running zoom simulations of the formation of different galaxy types, from Milky Way types, to early type galaxies and dwarfs, as well as environment (i.e., groups/clusters/isolated) it is possible to explicitly test the simulations against the wealth of observations available for GC populations and their relation to their host galaxy (see the review by Kruijssen~2014).

While still in early development, such simulations offer a huge promise to elucidate the co-formation/evolution of GCs and their host galaxies.

\subsection{Multiple Populations}
\label{sec:mps}

The lectures by Corinne Charbonnel in this volume nicely introduce the topic of multiple stellar populations within GCs.  Briefly, it is now clear that most GCs display star-to-star abundance variations, with only some elements (such as O, C, N, Na, Al, He) varying while others (Ca, Fe) remaining constant within the observational errors.  It is these abundance variations that cause the complex CMDs observed in GCs, when the appropriate filters are used (e.g., Piotto et al.~2015).  It should be emphasised that age differences or metallicity differences between the stars (stellar populations) are not the underlying cause of the features observed in high-resolution CMDs.  Age spreads of a few 10s of Myr may exist within the clusters, as we simply do not have that kind of age resolution at the ages of most GCs ($\sim10$~Gyr and above).  Hence, the observations are consistent with no age spreads, although age spreads of up to $\sim200-300$~Myr could be present within the uncertainties (Nardiello et al.~2015).

While it remains to be shown that YMCs and GCs share the same formation mechanism, none of the scenarios put forward to explain the origin of multiple populations (MPs) invokes any special mechanism that would only apply to GCs and not to YMCs.  The scenarios either implicitly or explicitly suggest that we can use YMCs, those with the expected properties of proto-GCs, to search for the predicted properties of multiple populations when GCs were of that age, i.e., multiple bursts of star-formation, large gas reservoirs, long proto-planetary disc lifetimes (e.g., Sollima et al.~2013).  There have been a number of works in the past $\sim5$~years that can be used to place tight constraints on the proposed scenarios.

For a recent compilation of the properties of clusters with and without multiple populations, see Krause et al.~(2016).  Below we highlight some tests that have been undertaken to use YMC properties to test models for the origin of multiple populations within GCs

\subsubsection{Testing the AGB Scenario}

In this scenario, a 2nd generation of stars forms out of a combination of stellar ejecta processed through high mass AGB stars and primordial material left over from the formation of the 1st generation.  There are two clear predictions of this scenario that can be tested using YMCs; the first is that clusters in the age range of $30-200$~Myr should be undergoing the 2nd generation of star-formation.  Secondly, in order to form the 2nd generation stars, these clusters should be gas rich, i.e. contain the gas/dust necessary to form the 2nd generation stars.

Peacock et al.~(2013) and Bastian et al.~(2013b) searched for evidence of ongoing star-formation within YMCs in the appropriate age range (and with masses extending from $10^4 - 10^8$~\msun) by looking for line emission (i.e. H$\beta$ or $[O\texttt{III}]$) in the integrated light of the clusters.  No clusters were found showing evidence of ongoing secondary star-formation.  Similarly, Larsen et al.~(2011) studied the CMDs of six nearby extragalactic YMCs ($10^5 - 10^6$~\msun) and did not find evidence of ongoing star-formation within them.  Niederhofer et al.~(2015) studied the resolved CMDs of slightly lower mass (few times $10^4 - 10^5$~\msun) clusters in the LMC and also did not find evidence for multiple bursts or extended SFHs predicted by the AGB scenario.

Cabrera-Ziri et al.~(2014) extended this type of study by estimating the star-formation history of a $100$~Myr, $10^7$~\msun\ cluster, NGC~34-1, using a high-S/N optical integrated light spectrum.  The cluster was found to be best reproduced by a single stellar population.  Multiple bursts, as predicted in the AGB scenario were found to be inconsistent with the observations.

Longmore (2015) showed that in the AGB scenario, at least in the simulations of D'Ercole et al.~(2008), the large reservoir of gas and dust that must be present within the cluster in order to form the second generation will lead to extremely high extinction values in the cluster centres.  The cluster outskirts will be largely unaffected (due to the strong central concentration of the gas/dust), so the luminosity profiles of the clusters should show strong dips in their centres.  This is not consistent with observations of YMCs.  Additionally, all or most YMCs should appear severely reddened in their integrated colours, whereas many YMCs with ages between 30-200~Myr are consistent with no internal reddening (e.g., Bastian et al.~2013b)

Cabrera-Ziri et al.~(2015) searched for gas/dust within 3 YMCs in the Antennae galaxies using ALMA, and were able to place upper limits of $1-10$\% of the stellar mass is present of gas/dust, in conflict with AGB model predictions.

To summarise, there is no evidence for multiple bursts of star-formation within YMCs (with ages between 10-1000~Myr) or for the retention of gas/dust required to form a 2nd generation (they appear to be gas free after 2-3 Myr, and no cluster older than this has been found with significant amounts of gas within it).  YMCs appear to be extremely efficient at expelling any gas within them, and there is no obvious reason to think that the same wouldn't be true for the proto-GCs.

\subsubsection{Testing the FRMS Scenario}

In the Fast Rotating Massive Star Scenario, as explored in Krause et al.~(2013), the proto-GC is unable to expel large amounts of gas/dust left over from the formation of the first generation of stars, for the first 20-30~Myr.  This allows time for a 2nd generation of stars to form in the excretion discs around rapidly rotating stars, and primordial material is also accreted onto the disc (in order to match the observed chemistry and to minimise the mass budget problem).

The predictions of this scenario have been tested by Bastian et al.~(2014), who searched for YMCs with ages less than $\sim20$~Myr that were gas free.  If YMCs are able to expel their gas on short timescales this would be in contradiction to the Krause et al. predictions, and would make it difficult to have enough time to form a 2nd generation within the allowed time.

Bastian et al.~(2014) found a number of massive young clusters ($>10^6$~\msun) that are gas free from a very young age ($2-3$~Myr).  Many of the YMCs have blown bubbles around them into the surrounding ISM, which can be used to estimate the time when the cluster was able to expel any remaining gas within it.  The authors found that, independent of metallicity, YMCs were able to expel their gas within $<2-3$~Myr after their formation.

Hollyhead et al.~(2015) expanded this study to look at lower mass clusters ($10^4 - 10^5$~\msun) in the nearby spiral galaxy M83.  They found a similar timescale, suggesting that independent of cluster mass, YMCs are extremely efficient at removing any gas within them, over very short timescales (see Fig.~\ref{fig:hollyhead}).

Another way to test the FRMS scenario, is to see whether stars within YMCs are in fact rotating very rapidly, i.e. near their breakup velocity.  An implicit assumption of this scenario is that stars in low density environments are not rotating near their critical velocity, while stars in massive clusters are, or else we would see the same chemical anomalies in the field.  In the recent VLT-FLAMES Tarantula Survey (VFTS - Evans et al.~2011), the rotation rate of hundreds of O and B stars in the massive YMC R136 and the surrounding association were measured.  Contrary to the expectations of the FRMS scenario, most of the stars in the region were slow rotators ($80$\% rotate at $<20$\% of the breakup velocity) and the fastest rotators were found outside R136 (Ram{\'{\i}}rez-Agudelo et al.~(2013))

\subsubsection{Testing the EDA Scenario}

Bastian et al.~(2013a) put forward a scenario for the formation of multiple populations in GCs that did not invoke multiple epochs of star-formation.  Instead, they used the discs around low-mass stars to act as nets, sweeping up enriched material ejecta from rotating massive stars or interacting high mass binaries, which eventually accretes onto the low mass star.  This is known as the Early Disc Accretion scenario, and was largely constructed with the constraints of YMCs in mind.

One major assumption of this scenario is that the discs around low-mass pre-main sequence stars can survive for $\sim10$~Myr within the dense environment of a proto-GC.  However, a number of effects can destroy such discs, namely interactions with passing stars (e.g., de Juan Ovelar et al.~2012) or ionising radiation from nearby massive stars (e.g., Scally \& Clarke~2001).  The fraction of stars with discs in nearby YMCs, such as NGC~3603, does appear to be much lower than required in the EDA scenario (e.g., Stolte et al.~2015).  Hence, it appears unlikely that this scenario can operate efficiently enough to explain the multiple populations observed in GCs.

\subsection{Mass is Not the Parameter Controlling Multiple Populations}

It is often stated that cluster mass is the parameter which controls whether or not a globular cluster hosts multiple populations or not.  This idea comes from the general dichotomy between globular clusters (which do host MPs) and open clusters (which do not host MPs - e.g., Bragaglia et al. 2012).  While there is not clear evidence for mass being the critical parameter, it has proven to be a stubborn view that has not gone away.

Krause et al.~(2016) compile a large list of clusters with and without MPs, as well as their properties.  There are a number of GCs with present day masses below $10^5$~\msun\ that host MPs (e.g., NGC~6362 - Dalessandro et al.~2014).  However, there are a number of young/intermediate age clusters that do not show evidence of MPs that have masses in excess of $10^5$~\msun\ (e.g., NGC 1806 - Mucciarelli et al.~2014; NGC~1846 - Mackey et al., in prep.).  These two clusters are both $\sim1.5$~Gyr old, and live in a relatively weak tidal environment, so they are not expected to lose much mass between the present day and when they will reach the age of current GCs.

Hollyhead et al. (in prep.) are extending this type of analysis to older ($6-8$~Gyr) Gyr clusters in the SMC, where the comparison between the present day GC and the younger clusters' masses is even more straightforward.

\begin{figure}
\centering
\includegraphics[width=10cm]{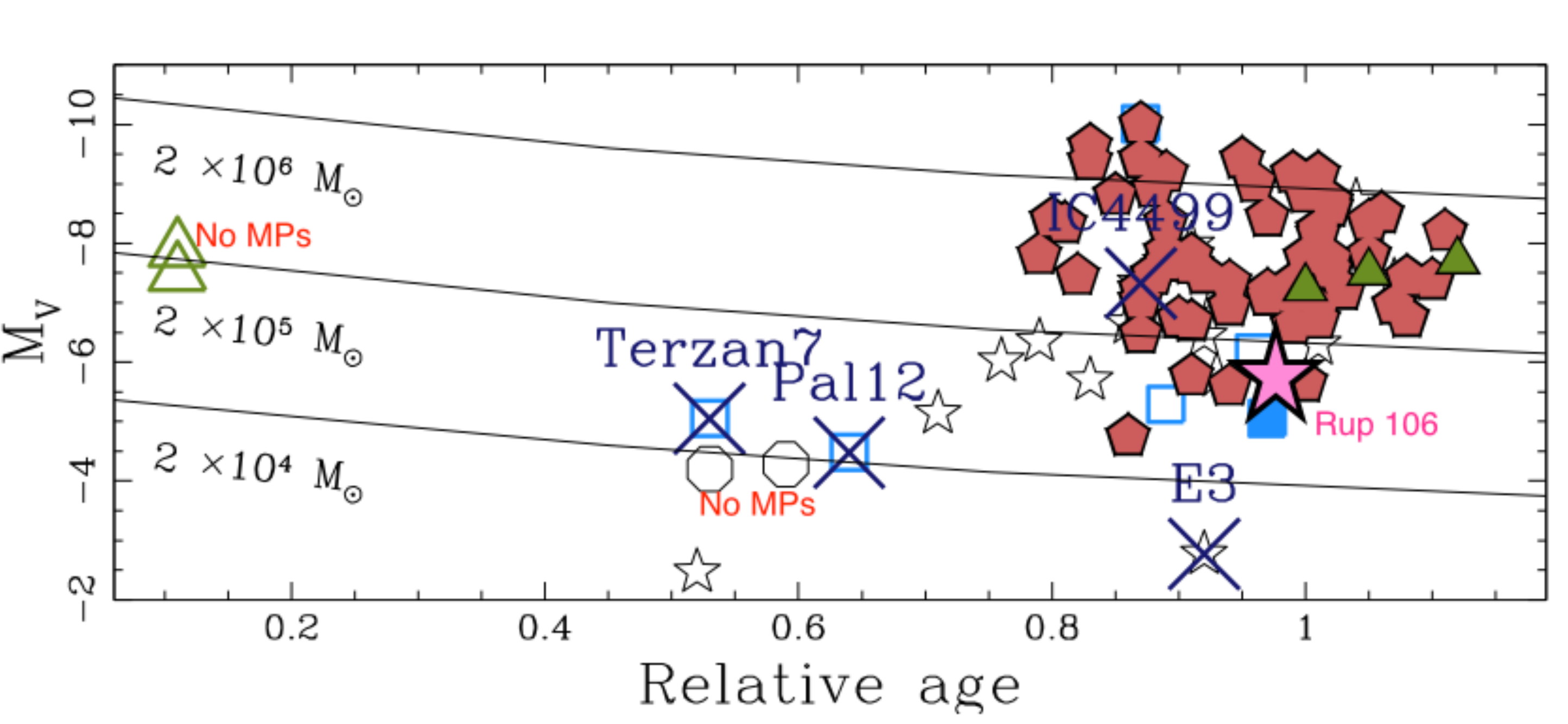}
\caption{Relative age  vs. absolute magnitude M$_{V}$ for globular and open clusters. Filled red symbols are GCs in the Milky Way (MW) and filled green triangles are GCs in the LMC; where MPs have been detected. Crossed denote GCs where studies have claimed that no MPs exist within the cluster.  Open star symbols and open squares mark MW and Sgr GCs, respectively, for which not enough data are available; open circles are the old open clusters Ber~39 and NGC6791 that do not show the Na inhomogeneities. Green empty triangles denote the intermediate age massive LMC clusters NGC~1806 and NGC~1846 that also do not host MPs (Mucciarelli et al.~2014; Mackey et al. in prep).   Blue crosses indicate our GCs that have been suggested to not host MPs and the filled star shows the position of Ruprecht~106 which does not host MPs and has been studied in sufficient detail to detect them (Villanova et al.~2013). Superimposed are lines of constant mass (adapted from Bragaglia et al. 2012). Figure courtesy of Carmela Lardo.}
\label{fig:no_mps}
\end{figure}

Fig.~\ref{fig:no_mps} shows the current status of surveys for MPs in massive clusters.  As was found in Krause et al.~(2016), there is no clear cluster parameter that controls whether it will host MPs or not.  However, it is clear that a simple mass cut is not consistent with the data\footnote{While it is also common to assume that GCs lost substantial fractions of their mass (losing more than $90-95$\% of their initial masses) in order to not violate the mass budget problem, this is inconsistent with the observed properties of the Galactic GC system (e.g., Bastian \& Lardo~2015) and is contradictory to dynamical expectations (e.g., Kruijssen~2015).}.

Further studies testing the extremes of the GC and nearby young/intermediate age cluster populations are necessary to see which property, if any, controls whether a cluster hosts MPs are not.  Is it age, i.e. some cosmological effect?  Birth location/environment?  

\section{Summary}

These lectures covered the main properties of YMCs and YMC populations in nearby galaxies.  While YMCs tend to be relatively easy to find and study in high-resolution imaging, a number of selection effects, biases, and loose terminology (i.e., defining what is meant by a ``cluster") have caused a lot of confusion within the field. When studying populations, a very important tool in any observers kit is the ability to create synthetic populations (i.e., with a given age distribution, sampled from a mass function, and converted into observational colour/magnitude space through conversion using SSP models).  This allows one to check whether observational effects are limiting the analysis, whether the sample contains enough clusters for robust conclusions, and potentially able to suggest complimentary ways to test the results.

There is now good agreement between the observed age distribution of clusters within normal galaxies and that expected from theoretical models that include known disruption effects (two-body relaxation, tidal effects, encounters with GMCs).  Additionally, studies have begun to reveal how the host environment affects the type of clusters that are formed (e.g., size-of-sample effects) and even the fraction of stars that form in clusters.  Putting these two pieces together we are now in a position to develop relatively detailed models for the formation and evolution of full cluster populations, from nearby galaxies to the ancient globular cluster populations that formed at high redshift.  This in turn offers  significant potential to use globular clusters to trace the formation and evolution of galaxies.

The proximity of YMCs in the nearby universe means that we can study their formation and stellar populations in much more detail that we can ever hope to in globular clusters.  Hence, YMCs have proven extremely useful in testing theories for GC formation, and in particular, the formation of multiple populations within them.  At the moment, none of the proposed scenarios for the multiple populations observed in GCs is consistent with observations of YMCs.  Individual YMCs appear to be gas free within $2-3$~Myr after formation and are extremely efficient at removing any gas within them (i.e. gas ejecta due to stellar evolution) for their entire subsequent evolution.  This is likely the reason why there is no compelling evidence for significant age spreads within YMCs, simply because they cannot retain the gas necessary to undergo multiple generations of star-formation.

There are no obvious properties of YMCs that show that they are distinct objects from GCs, but on the other hand there is also no result that shows unambiguously that they are indeed the same objects, merely separated by a Hubble time of evolution.  One major step in that direction would be if the characteristic property of GCs, i.e., multiple populations, were to be discovered, or ruled out, in YMCs.  This is going to be a particularly difficult measurement to make, and studies to date have had to make assumptions in their comparisons, i.e., comparing different stellar mass ranges in the YMCs (often $>15$~\msun) to GCs ($<0.8$~\msun), c.f. Cabrera-Ziri et al.~(2016b).  Hence, this promises to be a rich avenue for future work.


\end{document}